\documentclass[12pt]{article}
\usepackage{amsmath,amsfonts,graphicx}
\usepackage[T2A]{fontenc}
\usepackage[cp1251]{inputenc}
\usepackage[english]{babel}
\newcommand{\be}{\begin{equation}}
\newcommand{\ee}{\end{equation}}
\newcommand{\ba}{\begin{eqnarray}}
\newcommand{\ea}{\end{eqnarray}}
\numberwithin{equation}{section}
\makeatletter \@addtoreset{equation}{section} \makeatother

\begin{document}
\title{\Large \bf{On the effective lagrangian in
spinor\linebreak electrodynamics with added violation\linebreak
 of Lorentz and CPT symmetries}}

\author{Yurii A. Sitenko\thanks{E-mail:
yusitenko@bitp.kiev.ua}\\
\it \small Bogolyubov Institute for Theoretical Physics, National
\it \small  Academy of Sciences, 03143 Kyiv, Ukraine\\
 \it\small and \\
 \it\small Institute for Theoretical Physics, University of
 Berne,
 \it\small CH-3012 Berne, Switzerland\\
 \phantom{11111111111}\\
 Ksenya Yu. Rulik
\thanks{E-mail: xenia@univ.kiev.ua}\\
\it \small Bogolyubov Institute for Theoretical Physics, National
\it \small Academy of Sciences, 03143 Kyiv , Ukraine\\
}
\date{}
\maketitle

\centerline{\textbf{Abstract}}
 We consider quantum electrodynamics with additional coupling of
spinor fields to the space-time independent axial vector violating
both Lorentz and CPT symmetries. The Fock-Schwinger proper time
method is used to calculate the one-loop effective action up to
the second order in the axial vector and to all orders in the
space-time independent electromagnetic field strength. We find
that the Chern-Simons term is not radiatively induced and that the
effective action is CPT invariant in the given approximation.
\renewcommand{\theequation}{\arabic{section}.\arabic{equation}}

\begin{flushleft}

\end{flushleft}

\section{Introduction}

\hspace{0.7cm}Although conservation of the Lorentz and CPT
symmetries belongs to the fundamental laws of nature, various
extensions of quantum electrodynamics and, more generally, the
standard model with tiny violation of these symmetries have
generated current interest in the last decade \cite{Ca, Co1, CG,
Ja1,Kost}. In gauge vector sector of quantum electrodynamics a
plausible extension is achieved \cite{Ca} by adding a Chern-Simons
term \cite{Jao} to the conventional Maxwell term in the lagrangian
 \be \mathcal{L}(A,k) = \frac{1}{4e^2}SpF^2 - k\tilde{F}A, \ee
  where $A^{\mu}$ is the electromagnetic potential, $F_{\mu\nu} = \partial_{\mu}A_{\nu} -
  \partial_{\nu}A_{\mu}$ is the field strength tensor, $\tilde{F}^{\mu\nu} =
  \frac{1}{2}\varepsilon^{\mu\nu\alpha\beta}F_{\alpha\beta}$ is its
  dual, $Sp$ is the trace over the Lorentz indices: $SpMN =
  M_{\mu\nu}N^{\nu\mu}$, and metric of the Minkowski space is
  chosen as $g_{\mu\nu} = diag(1, -1, -1, -1)$. Second term in the
  right hand side of Eq.(1.1) is obviously non-gauge invariant,
  however, if vector $k^{\mu}$ is space-time independent, then the
  integral of the second term over the whole space-time is gauge
  invariant. Consequently, the action and equations of motion of
  the theory with lagrangian (1.1) are  gauge invariant. Position
  independent vector $k^{\mu}$ selects a preferred direction in
  space-time, thus violating both the Lorentz and CPT symmetries.
  Observation of distant galaxies puts a stringent bound on the
  value of $k^{\mu}$: it should effectively vanish \cite{Ja1, Go}.

  An obvious extension of the spinor sector of quantum
  electrodynamics is
 \be \mathcal{L}(\psi, \bar{\psi}, A, b) = \overline{\psi}(\imath\hat{\partial} -
\hat{A} + \hat{b}\gamma^{5} - m)\psi,\ee
 where
$\hat{\partial}= \gamma^{\mu}\partial_{\mu}$,
$[\gamma^{\mu},\gamma^{\nu}]_{+}=2g^{\mu\nu}$ and
$\gamma^5=-\imath\gamma^0\gamma^1\gamma^2\gamma^3$. If vector
$b^{\mu}$ is space-time independent, then a natural question
arises, whether a Chern-Simons term can be radiatively induced as
a result of interaction of quantized spinor fields in the theory
with lagrangian (1.2). Different answers  to this question have
been obtained,  which can be summarized as follows. Perturbative
(and even nonperturbative but based on the Feynman diagram
technique) calculations yield the induced Chern-Simons term with
fixed coefficient, but the value of this coefficient differs
depending on  a concrete calculation scheme \cite{Ja2, WC, PV,
JC1, And}. This discrepancy is analyzed further, and the claim is
made that the Chern-Simons term is induced with finite but
intrinsically ambiguous coefficient \cite{Ja3, Pe, Ba}. There are
also nonperturbative (nondiagrammatic or functional) approaches
which yield either fixed or ambiguous values of the coefficient
before the induced Chern-Simons term \cite{JC2, MC, JC3, Si1}. At
last there are rather diverse arguments that the Chern-Simons term
is not radiatively induced \cite{Co1, CG, Bo, Ada}.

To shed more light on this problem, we shall compute the effective
action of the theory with lagrangian (1.2) in the approximation
keeping all orders of position independent electromagnetic field
strength $F_{\mu\nu}$  and up to the second order in position
independent vector $b^{\mu}$. We use the Fock-Schwinger
\cite{Fo,Sch} proper time method and find out that, indeed, the
Chern-Simons term is not induced. Moreover, the linear in
$b^{\mu}$ terms are absent at all, and the effective action is
parity invariant.

\section{Effective action and its regularization}

\hspace{0.7cm}The effective action is obtained by integrating out the fermionic
degrees of freedom in the theory with lagrangian (1.2):
 \ba && \nonumber \Gamma(A,b) =  -\imath\ln\int
d\bar{\psi}d\psi\exp\left[\imath \int d^{4}x\mathcal{L}(\psi,
\bar{\psi}, A, b)\right] = \\ && = -\imath\ln Det(\imath
\hat{\partial} - \hat{A} + \hat{b}\gamma^5 - m) = \int d^4
x\mathcal{L}^{eff}(A, b), \ea
 where
 \be \mathcal{L}^{eff}(A, b) = -\imath tr<x|\ln(\imath\hat{\partial} - \hat{A} + \hat{b}\gamma^{5}
- m)|x> \ee
  is the effective lagrangian, and the trace over spinor indices is denoted by $tr$.
In the most general case (i.e. for arbitrary space-time dependent
vectors $A^{\mu}(x)$ and $b^{\mu}(x)$) the effective action can be
represented as a sum of two terms
  \be \Gamma(A,b) =
-\frac{\imath}{2} Tr\ln\mathcal{H} -
Tr\arctan[(\hat{\partial}+\imath\hat{A}-\imath\hat{b}\gamma^5)m^{-1}],
\ee
 where
 \be \mathcal{H} =
-(\imath\hat{\partial}-\hat{A}+\hat{b}\gamma^5)^2+m^2, \ee
 and $Tr$ is the trace of the differential operator in functional
 space: $Tr U = \int d^4 x \\tr<x|U|x>$.
 Since the trace of odd number of $\gamma$-matrices vanishes, one gets
relation
 \be \frac{\delta}{\delta
A_{\mu}(x)}Tr\arctan[(\hat{\partial}+\imath\hat{A}-\imath\hat{b}\gamma^5)m^{-1}]=
\frac{\delta}{\delta
b_{\mu}(x)}Tr\arctan[(\hat{\partial}+\imath\hat{A}-\imath\hat{b}\gamma^5)m^{-1}]=0,
\ee
 and, therefore, the second term in the right-hand side of
Eq.(2.3) can be neglected as inessential. As to the first  term in
the right-hand side of Eq.(2.3), it can be related to the zeta
function of operator $\mathcal{H}$ \cite{Sal, Dow, Haw}
 \be
-\frac{\imath}{2}Tr\ln\mathcal{H}=\frac{\imath}{2}\left(\frac{d}{dz}Tr\mathcal{H}^{-z}\right)|_{z=0}.\ee
 Using integral representation for the zeta function
 \be Tr\mathcal{H}^{-z}=
\frac{1}{\Gamma(z)}\int^{\infty}_{0}d\tau\tau^{z-1}Tr
e^{-\tau\mathcal{H}}, \hspace{1cm} Re z>0, \ee where $\Gamma(z)$
is the Euler gamma function, one gets integral representation for
the effective action
 \be \Gamma(A,b)=
\frac{\imath}{2}\int^{\infty}_{0}\frac{d\tau}{\tau}Tr
e^{-\tau\mathcal{H}}.\ee

 Taking functional derivatives of
Eq.(2.3), let us define vector current
 \be J^{\mu}(x) \equiv
-\frac{\delta\Gamma(A,b)}{\delta A_{\mu}(x)} = \imath
tr\gamma^{\mu}G(x,x), \ee and axial-vector current
 \be J^{\mu5}(x)
\equiv \frac{\delta\Gamma(A,b)}{\delta b_{\mu}(x)} = \imath
tr\gamma^{\mu}\gamma^5G(x,x), \ee where
 \be
G(x,y)=<x|(\imath\hat{\partial}-\hat{A}+\hat{b}\gamma^5+m)\mathcal{H}^{-1}|y>,
\ee is the Green's function. One can write integral representation
for the latter:
 \be
G(x,y)=\int^{\infty}_{0}d\tau<x|(\imath\hat{\partial}-\hat{A}+\hat{b}\gamma^5+m)e^{-\tau\mathcal{H}}|y>.\ee
 In the case when $b_{\mu}$ is space-time independent, the
effective action can be presented in the form
 \be \Gamma(A,b) =
\Gamma(A,0)+ \int^{1}_{0}du\int d^4x b_{\mu}J^{\mu5}(x;u) ,\ee
where $J^{\mu5}(x;u)$ is the axial-vector current with $ub$
substituted for $b$.

 It should be emphasized that
most of the above relations are purely formal, since they suffer
from ultraviolet divergencies. For instance, Green's function
(2.11) is well-defined at $x\neq y$, and diverges at $x\rightarrow
y$. To regularize the divergence, one introduces a cut off at the
lower limit of the integral in representation (2.12). In this way
one gets a regularized definition of currents (2.9) and (2.10),
which, after appropriate integration, yield the regularized
expression for the effective action.

In the present paper we restrict ourselves to the case of
space-independent field tensor $F^{\mu\nu}$ and vector $b^{\mu}$.
Then operator $\mathcal{H}$ (2.4) takes form \be \mathcal{H} =
-\pi^{\mu}\pi_{\mu} - 2\imath\gamma^{5}\sigma^{\mu\nu}
b_{\mu}\pi_{\nu} + \frac{1}{2}F_{\mu \nu}\sigma^{\mu \nu} +
b^{\mu}b_{\mu} + m^{2}, \ee where
 \be \pi_{\mu}=\imath\partial_{\mu} -
A_{\mu},\hspace{0.5cm}
\sigma^{\mu\nu}=\frac{\imath}{2}[\gamma^{\mu},\gamma^{\nu}]_{-}.\ee
Our aim is
to find currents (2.9) and (2.10) and then effective lagrangian (2.2).

\section{Proper time method}

 \hspace{0.7cm}Rotating the integration path in Eq.(2.12) by angle $\pi/2$ in the
anticlockwise direction (i.e. substituting $\tau$ by $\imath s$),
we present the Green's function in the form

\be G(x,y) = \imath\int^{\infty}_{0}ds<x|(\hat{\pi} + \hat{b}\gamma^{5} +
m)e^{-\imath s\mathcal{H}}|y>, \ee
where it is implied that the mass squared in
$\mathcal{H}$ entails a small negative imaginary part, $m^2\to m^2 -
i\epsilon$. The idea of the proper time method  of Fock \cite{Fo} and Schwinger
\cite{Sch} is to treat operator $\mathcal{H}$ as a Hamilton operator that
governs evolution in "time" $s$ of a hypothetical quantum mechanical system.
Then transition amplitude (matrix element of evolution operator $\exp (-\imath
s\mathcal{H})$)
 \be <x|e^{-\imath
s\mathcal{H}}|y> \equiv <x(s)|y(0)>,\ee where
 \be |y(0)> = |y>, \hspace{1cm}
|x(s)> = e^{\imath s\mathcal{H}}|x>, \ee satisfies evolution
equation
 \be \imath\partial_{s}<x(s)|y(0)> =
<x(s)|\mathcal{H}|y(0)>, \ee with boundary conditions
 \be
\lim_{s\rightarrow 0}<x(s)|y(0)> = \delta (x-y), \hspace{1cm}
\lim_{s\rightarrow \infty}<x(s)|y(0)> = 0. \ee
 Commutation
relations \ba \nonumber && [x^{\mu},\pi^{\nu}]_{-} = -\imath
g^{\mu \nu}, \hspace{1cm} [\pi^{\mu},\pi^{\nu}]_{-} = -\imath
F^{\mu \nu},
\\&& [\sigma^{\mu\nu},\sigma^{\omega\rho}]_{-} =
2\imath(\sigma^{\mu\rho}g^{\nu\omega} -
\sigma^{\nu\rho}g^{\mu\omega} - \sigma^{\mu\omega}g^{\nu\rho} +
\sigma^{\nu\omega}g^{\mu\rho}), \\ &&\nonumber
[x^{\mu},x^{\nu}]_{-} = [x^{\mu},\sigma^{\omega\rho}]_{-} =
[\pi^{\mu},\sigma^{\omega\rho}]_{-} = 0, \ea are invariant under
unitary transformation
 \ba X^{\mu}(s) &=& e^{\imath
s\mathcal{H}}X^{\mu}(0)e^{-\imath s\mathcal{H}}, \hspace{1cm}
\Pi^{\mu}(s) = e^{\imath s\mathcal{H}}\Pi^{\mu}(0)e^{-\imath
s\mathcal{H}},\\ \nonumber \Sigma^{\mu\nu}(s) &=& e^{\imath
s\mathcal{H}}\Sigma^{\mu\nu}(0)e^{-\imath s\mathcal{H}},\ea where
 \be X^{\mu}(0) = x^{\mu}, \hspace{1cm}\Pi^{\mu}(0) = \pi^{\mu},
\hspace{1cm} \Sigma^{\mu\nu}(0) = \sigma^{\mu\nu},\ee can be
regarded as quantum mechanical observable operators in the
Schrodinger representation, while $X^{\mu}(s), \Pi^{\mu}(s),
\Sigma^{\mu\nu}(s)$ as those in the Heisenberg representation. The
latter satisfy evolution equations:
 \be  \dot{X}^{\mu}(s) = \imath
[\mathcal{H},X^{\mu}(s)]_{-} ,\hspace{0.5cm} \dot{\Pi}^{\mu}(s) =
\imath [\mathcal{H},\Pi^{\mu}(s)]_{-} , \hspace{0.5cm}
\dot{\Sigma}^{\mu\nu}(s) = \imath
[\mathcal{H},\Sigma^{\mu\nu}(s)]_{-}.\ee
 Using the explicit form
of Hamilton operator $\mathcal{H}$ and commutation relations
(3.6), one can compute commutators in the right hand sides of
Eq.(3.9), and then solve this system of equations. Using the
solution, one can compute the matrix element in the right hand
side of Eq.(3.4), and then solve this equation and find transition
amplitude (3.2). The consistency relations must hold:
 \ba
\nonumber <x(s)|\Pi_{\mu}(s)|y(0)> &=&
[\imath\partial^{(x)}_{\mu} - A_{\mu}(x)]<x(s)|y(0)>,\\
\nonumber <x(s)|\Pi_{\mu}(0)|y(0)> &=&
[-\imath\partial^{(y)}_{\mu} - A_{\mu}(y)]<x(s)|y(0)>,\\
<x(s)|X^{\mu}(s)|y(0)> &=& x^{\mu}<x(s)|y(0)>,\\ \nonumber
<x(s)|X^{\mu}(0)|y(0)> &=& y^{\mu}<x(s)|y(0)>,\\ \nonumber
<x(s)|\Sigma^{\mu\nu}(s)|y(0)> &=& \sigma^{\mu\nu}<x(s)|y(0)>,\\
\nonumber <x(s)|\Sigma^{\mu\nu}(0)|y(0)> &=&
<x(s)|y(0)>\sigma^{\mu\nu}. \ea

Let us find the transition amplitude in the case of $\mathcal{H}$
given by Eq.(2.14) with constant uniform electromagnetic field
strength. System of evolution equations (3.9) takes form
 \ba
\nonumber \dot{X}^{\mu}(s) &=&  2\Pi^{\mu}(s) - 2\imath
\Sigma^{\mu\nu}(s)b_{\nu}\gamma^{5}, \hspace{0.5cm} \dot{\Pi}_{\mu}(s) =
2F_{\mu\nu}\Pi^{\nu}(s) -2 \imath F_{\mu\nu}
\Sigma^{\nu\omega}(s)b_{\omega}\gamma^{5}, \\
\dot{\Sigma}^{\mu\nu}(s) &=&
2[F^{\mu\omega}g_{\omega\rho}\Sigma^{\rho\nu}(s) -
\Sigma^{\mu\omega}(s)g_{\omega\rho}F^{\rho\nu}] +
4\imath\{[\Pi^{\mu}(s)b_{\omega} -
b^{\mu}\Pi_{\omega}(s)]\Sigma^{\omega\nu}(s) -\\ &&\nonumber -
\Sigma^{\mu\omega}(s)[\Pi_{\omega}(s)b^{\nu} -
b_{\omega}\Pi^{\nu}(s)]\}\gamma^{5}. \ea The system is solved in
the linear in $b$ approximation yielding, in obvious matrix
notations,
 \ba \nonumber &&X(s) - X(0) =
2e^{Fs}\frac{\sinh(Fs)}{F}\Pi(0) - 2\imath
e^{2Fs}\Sigma(0)\frac{\sinh(Fs)}{F}e^{-Fs}b\gamma^{5},
\\ && \Pi(s) = e^{2Fs}\Pi(0) - 2\imath
e^{2Fs}F\Sigma(0)\frac{\sinh(Fs)}{F}e^{-Fs}b\gamma^{5}, \\
\nonumber && \Sigma(s) = e^{2Fs}\Sigma(0)e^{-2Fs} + 4\imath
e^{2Fs}\{[\Pi(0)be^{Fs}\frac{\sinh(Fs)}{F} -
\frac{\sinh(Fs)}{F}e^{-Fs}b\Pi(0)]\Sigma(0) -\\ \nonumber &-&
\Sigma(0)[\Pi(0)be^{Fs}\frac{\sinh(Fs)}{F} -
\frac{\sinh(Fs)}{F}e^{-Fs}b\Pi(0)]\}e^{-2Fs}\gamma^{5}. \ea
 Using two last
relations in Eq.(3.12), we get
 \ba \mathcal{H} &=& -\Pi^{2}(s) +
2\imath\gamma^{5}\Pi(s)\Sigma(s)b - \frac{1}{2}SpF\Sigma(s) +
m^{2} = -\Pi^{2}(0)  + \\ \nonumber &+&
2\imath\gamma^{5}\Pi(0)\Sigma(0)b -  \frac{1}{2}Sp F\Sigma(0) +
m^{2}. \ea
 The last equation states that the
Hamilton operator in the Heisenberg representation coincides with
that in the Schrodinger representation being independent of $s$,
as it should be. However, the matrix element of $\mathcal{H}$ in
the right hand side of Eq.(3.4) is s-dependent, and, to find this
dependence, one has to express operator (3.13) through operators
$X(s), X(0)$ and either $\Sigma(s)$ or $\Sigma(0)$. Using the
first relation in Eq.(3.12), we get
 \ba \Pi(0) &=& \frac{e^{-Fs}F}{2\sinh(Fs)}[X(s) - X(0)] +
\imath
e^{Fs}\frac{F}{\sinh(Fs)}\Sigma(0)\frac{\sinh(Fs)}{F}e^{-Fs}b\gamma^{5},\\
\nonumber \Pi(s) &=& \frac{e^{Fs}F}{2\sinh(Fs)}[X(s) - X(0)] +
\imath
e^{-Fs}\frac{F}{\sinh(Fs)}\Sigma(s)\frac{\sinh(Fs)}{F}e^{Fs}b\gamma^{5},
\ea and, consequently,
 \ba \Pi^{2}(s) &=& \frac{1}{4}[X(s) -
X(0)]\frac{F^{2}}{\sinh^{2}(Fs)}[X(s) - X(0)] + \\ \nonumber &+&
\imath [X(s) -
X(0)]\frac{e^{-2Fs}F^{2}}{\sinh^{2}(Fs)}\Sigma(s)\frac{\sinh(Fs)}{F}e^{Fs}b\gamma^{5}.
\ea Using commutational relation
 \be [X_{\mu}(s),X_{\nu}(0)]_{-} =
\imath\left(\frac{e^{Fs}\sinh(Fs)}{F}\right)_{\mu\nu}, \ee
 and adding relevant terms to $-\Pi^{2}(s)$, we get $\mathcal{H}$ (3.13)
as a proper-time-ordered function of X(s) and X(0)
 \begin{subequations}
 \ba \nonumber \mathcal{H} &=&
-\frac{1}{4}X(s)\frac{F^{2}}{\sinh^{2}(Fs)}X(s)
+\frac{1}{2}X(s)\frac{F^{2}}{\sinh^{2}(Fs)}X(0) -
\frac{1}{4}X(0)\frac{F^{2}}{\sinh^{2}(Fs)}X(0) - \\
 &-& \frac{\imath}{2}Sp F\coth(Fs) -
\frac{1}{2}Sp F\Sigma(s) + m^{2} + \\
\nonumber &+& \imath[X(s) -
X(0)]\frac{e^{-Fs}F}{\sinh(Fs)}\left[\Sigma(s) -
\frac{e^{-Fs}F}{\sinh(Fs)}\Sigma(s)\frac{\sinh(Fs)}{F}e^{Fs}\right]b\gamma^{5},
\ea or, alternatively

\ba \nonumber \mathcal{H} &=&
-\frac{1}{4}X(s)\frac{F^{2}}{\sinh^{2}(Fs)}X(s)
+\frac{1}{2}X(s)\frac{F^{2}}{\sinh^{2}(Fs)}X(0) -
\frac{1}{4}X(0)\frac{F^{2}}{\sinh^{2}(Fs)}X(0) -
\\ &-& \frac{\imath}{2}SpF\coth(Fs) -
\frac{1}{2}SpF\Sigma(0) +  m^{2} - \\
\nonumber &-& \imath\gamma^{5}b\left[\Sigma(0) -
\frac{e^{Fs}F}{\sinh(Fs)}\Sigma(0)\frac{\sinh(Fs)}{F}e^{-Fs}\right]
\frac{e^{-Fs}F}{\sinh(Fs)}[X(s) - X(0)],
 \ea
\end{subequations}
Using the last four relations in Eq.(3.10), we find two forms of
the matrix element of $\mathcal{H}$
 \be <x(s)|\mathcal{H}|y(0)> =
P^{(a)}(x,y;s)<x(s)|y(0)> = <x(s)|y(0)>P^{(b)}(x,y;s), \ee where
 \begin{subequations}
 \ba \nonumber &&  P^{(a)}(x,y;s) =
-\frac{1}{4}(x-y)\frac{F^{2}}{\sinh^{2}(Fs)}(x-y) -
\frac{\imath}{2}Sp F\coth(Fs) - \frac{1}{2}Sp F\sigma + m^2 +
\\  &+&  \imath(x-y)\frac{e^{-Fs}F}{\sinh(Fs)}\left[\sigma -
\frac{e^{-Fs}F}{\sinh(Fs)}\sigma\frac{\sinh(Fs)}{F}e^{Fs}\right]b\gamma^{5},
\ea and
 \ba \nonumber &&  P^{(b)}(x,y;s) =
-\frac{1}{4}(x-y)\frac{F^{2}}{\sinh^{2}(Fs)}(x-y) -
\frac{\imath}{2}Sp F\coth(Fs)  - \frac{1}{2}Sp F\sigma + m^{2}-
\\   &-& \imath\gamma^{5}b\left[\sigma -
e^{Fs}\frac{\sinh(Fs)}{F}\sigma\frac{Fe^{-Fs}}{\sinh(Fs)}
\right]\frac{e^{-Fs}F}{\sinh(Fs)}(x-y), \ea
\end{subequations}

Substituting Eq.(3.19) into the right hand side of evolution
equation (3.4), and solving the latter, we get two equivalent
expressions for the transition amplitude:
 \begin{subequations}
 \ba &&\nonumber <x(s)|y(0)> =
-\frac{\imath}{(4\pi)^{2}}\exp\left\{-\imath \int^{x}_{y}d\xi[A
+\frac{1}{2}F(\xi-y)]\right\}\frac{1}{s^{2}} \times \\ \nonumber
&\times& \exp\left[-\frac{\imath}{4}(x-y)F\coth(Fs)(x-y) -
\frac{1}{2}Sp\ln\frac{\sinh(Fs)}{Fs} - \imath sm^{2}\right] \times
\\
 &\times&
\exp\left[(x-y)e^{-Fs}\frac{F}{\sinh(Fs)}\sigma\frac{\sinh(Fs)}{F}e^{Fs}b\gamma^{5}
\right]\exp\left(\frac{\imath s}{2}Sp F\sigma\right), \ea
 and
 \ba &&\nonumber <x(s)|y(0)> =
-\frac{\imath}{(4\pi)^{2}}\exp\left\{-\imath \int^{x}_{y}d\xi[A
+\frac{1}{2}F(\xi-y)]\right\}\frac{1}{s^{2}}\times \\ \nonumber
&\times& \exp\left[-\frac{\imath}{4}(x-y)F\coth(Fs)(x-y) -
\frac{1}{2}Sp\ln\frac{\sinh(Fs)}{Fs} - \imath sm^{2}\right] \times
\\ &\times& \exp\left(\frac{\imath s}{2}SpF\sigma\right)
\exp\left[-\gamma^{5}be^{Fs}\frac{\sinh(Fs)}{F}
\sigma\frac{F}{\sinh(Fs)}e^{-Fs}(x-y)\right], \ea
\end{subequations}
 where the $b$ dependent exponential should be understood as
 expanded up to the first order in $b$.
 Also we get relations
 \begin{subequations}
 \ba \nonumber   \gamma^{\mu}<x(s)|\Pi_{\mu}(s)|y(0)> &=&
\left[\frac{1}{2} \gamma e^{Fs}\frac{F}{\sinh(Fs)}(x-y) + \imath
\gamma
e^{-Fs}\frac{F}{\sinh(Fs)}\sigma\frac{\sinh(Fs)}{F}e^{Fs}b\gamma^{5}\right]\times
\\ &\times& <x(s)|y(0)>, \ea
 and
\ba  \nonumber  <x(s)|\Pi_{\mu}(0)|y(0)>\gamma^{\mu} &=& <x
(s)|y(0)>\left[\frac{1}{2}(x-y)e^{Fs}\frac{F}{\sinh(Fs)}\gamma
-\right. \\ && \left. - \imath
\gamma^{5}be^{Fs}\frac{\sinh(Fs)}{F}\sigma\frac{F}{\sinh(Fs)}e^{-Fs}\gamma\right],
 \ea
\end{subequations}
 and one can verify that the first two relations in Eq.(3.10) are
valid. Consequently, we get two equivalent representations for
Green's function $G(x,y)$ which, after rotating the integration
path in Eq.(3.1) back $(s=-\imath\tau)$, take form
 \begin{subequations}
 \ba &&\nonumber G(x,y) =
\int^{\infty}_{0}d\tau\left[\frac{\imath}{2}\gamma e^{-\imath
F\tau}\frac{F}{\sin(F\tau)}(x-y) +  \imath\gamma e^{\imath
F\tau}\frac{F}{\sin(F\tau)}\sigma\frac{\sin(F\tau)}{F}e^{-\imath
F\tau}b\gamma^{5}
 + \right. \\ && \left. +  \gamma b\gamma^{5} + m\right] <x(-\imath\tau)|y(0)>, \ea
 and
 \ba &&\nonumber G(x,y) = \int^{\infty}_{0}d\tau<x(-\imath\tau)|y(0)>
\left[\frac{\imath}{2}(x-y) \frac{F}{\sinh(F\tau)}e^{-\imath
F\tau}\gamma - \right. \\&& \left. -\imath\gamma^{5}b e^{-\imath
F\tau}\frac{\sin(F\tau)}{F}\sigma\frac{F}{\sin(F\tau)}e^{\imath
F\tau}\gamma - \gamma^{5}b\gamma + m\right]. \ea
\end{subequations}

\section{Currents}

\hspace{0.7cm}Inserting either Eq.(3.20a) or (3.20b) into
Eqs.(3.22a) and (3.22b), taking limit $y\rightarrow x$ and
retaining terms which are not higher than first order in
$b^{\mu}$, we get two equivalent expressions for vector current
(2.9):
 \begin{subequations}
 \ba \nonumber J^{\mu} &=& -\frac{1}{(4\pi)^{2}}\int^{\infty}_{0}\frac{d\tau}{\tau^{2}}\exp[-\tau m^{2} -\frac{1}{2}Sp\ln\frac{\sin(F\tau)}{F\tau}
]\times\\ &\times& tr\gamma^5\gamma^{\mu}\gamma[\imath e^{\imath
F\tau}\frac{F}{\sin(F\tau)}\sigma\frac{\sin(F\tau)}{F}e^{-\imath
F\tau}+1]b \exp(\frac{\tau}{2}Sp\sigma F),\ea
 and
 \ba  \nonumber J'^{\mu} &=& \frac{1}{(4\pi)^{2}}\int^{\infty}_{0}\frac{d\tau}{\tau^{2}}\exp[-\tau m^{2} -\frac{1}{2}Sp\ln\frac{\sin(F\tau)}{F\tau}
]\times\\ &\times& tr\exp(\frac{\tau}{2}Sp\sigma F)b[\imath
e^{-\imath
F\tau}\frac{\sin(F\tau)}{F}\sigma\frac{F}{\sin(F\tau)}e^{\imath
F\tau}+1]\gamma\gamma^{\mu}\gamma^5,\ea
\end{subequations}
 and for axial-vector current (2.10):
 \begin{subequations}
 \ba    \nonumber J^{\mu5} &=&
\frac{1}{(4\pi)^{2}}\int^{\infty}_{0}\frac{d\tau}{\tau^{2}}\exp[-\tau
m^{2} -\frac{1}{2}Sp\ln\frac{\sin(F\tau)}{F\tau} ]\times\\
&\times& tr\gamma^{\mu}\gamma[\imath e^{\imath
F\tau}\frac{F}{\sin(F\tau)}\sigma\frac{\sin(F\tau)}{F}e^{-\imath
F\tau}+1]b\exp(\frac{\tau}{2}Sp\sigma F),\ea
 and
 \ba \nonumber J'^{\mu5} &=& \frac{1}{(4\pi)^{2}}\int^{\infty}_{0}\frac{d\tau}{\tau^{2}}\exp[-\tau m^{2} -\frac{1}{2}Sp\ln\frac{\sin(F\tau)}{F\tau}
]\times\\ &\times& tr\exp(\frac{\tau}{2}Sp\sigma F)b[\imath
e^{-\imath
F\tau}\frac{\sin(F\tau)}{F}\sigma\frac{F}{\sin(F\tau)}e^{\imath
F\tau}+1]\gamma\gamma^{\mu}.\ea
\end{subequations}
One can notice that
 \be (J^{'\mu})^{*}=J^{\mu}, \hspace{1cm}
 (J^{'\mu5})^{*}=J^{\mu5},\ee
 and, therefore, both vector and axial-vector currents have to be real, if representations
 (3.22a) and (3.22b) are indeed equivalent.

 In order to take traces over $\gamma$-matrices in
 Eqs.(4.1)--(4.2), one uses expansion of the exponential of the
 $\sigma$-matrix,
 \be \exp\left(\frac{\tau}{2}Sp\sigma F\right) = C_{1}(\tau)I + C_{2}(\tau)Sp\sigma F + C_{3}(\tau)\imath
\gamma^{5} + C_{4}(\tau)\imath \gamma^{5}Sp\sigma F, \ee
 and relations
 \be (\tilde{F}^2)_{\mu\nu}=(F^2)_{\mu\nu} -
 \frac{1}{2}g_{\mu\nu}SpF^2, \hspace{0.5cm}
 (F\tilde{F})_{\mu\nu}=(\tilde{F}F)_{\mu\nu}=\frac{1}{4}g_{\mu\nu}Sp(F\tilde{F}),
 \ee
 \be \nonumber
 \varepsilon^{\mu\nu\alpha\beta}(FK)_{\alpha\beta}=-(\tilde{F}K)^{\mu\nu} - (K\tilde{F})^{\mu\nu}+\tilde{F}^{\mu\nu}SpK,\ee
 where $K_{\mu\nu}$ is an arbitrary symmetric second-rank Lorentz
 tensor, and evident identities $tr\gamma^{5}\gamma^{\mu}\gamma^{\nu}\sigma^{\alpha\beta} =
-4\varepsilon^{\mu\nu\alpha\beta}$,
$tr\gamma^{\mu}\gamma^{\nu}\sigma^{\alpha\beta} =
-4\imath(g^{\mu\alpha}g^{\nu\beta} - g^{\mu\beta}g^{\nu\alpha})$.
 Thus, we get
  \be ReJ^{\mu}=\frac{1}{(2\pi)^2}\int_{0}^{\infty}d\tau e^{-\tau
  m^2}\omega^{\mu\nu}(\tau)b_{\nu}, \hspace{0.5cm} ImJ^{\mu}=\frac{1}{(2\pi)^2}\int_{0}^{\infty}d\tau e^{-\tau
  m^2}\rho^{\mu\nu}(\tau)b_{\nu}, \ee
   and
   \be ReJ^{\mu5}=\frac{1}{(2\pi)^2}\int_{0}^{\infty}d\tau e^{-\tau
  m^2}\omega^{\mu\nu}_{5}(\tau)b_{\nu}, \hspace{0.5cm} ImJ^{\mu5}=\frac{1}{(2\pi)^2}\int_{0}^{\infty}d\tau e^{-\tau
  m^2}\rho^{\mu\nu}_{5}(\tau)b_{\nu}, \ee
 where
  \ba \nonumber && \omega^{\mu\nu}(\tau) =
-C_{0}(\tau)\left\{C_{1}(\tau)\left(\tilde{F}\frac{\sin2F\tau}{F}\right)^{\mu\nu}
+ C_{2}(\tau)[4\tilde{F}^{\mu\nu}-
\left(\tilde{F}\frac{\sin2F\tau}{F}\right)^{\mu\nu}Sp F\cot F\tau]
+ \right.\\ &&  + C_{3}(\tau)[(\sin2F\tau)^{\mu\nu} -
\left(\frac{\sin^{2}F\tau}{F}\right)^{\mu\nu}Sp F\cot F\tau] -
C_{4}(\tau)[4(F\cos2F\tau)^{\mu\nu}- \\
\nonumber && \left. - (\sin2F\tau)^{\mu\nu}Sp F\cot F\tau +
2\left(\frac{\sin^{2}F\tau}{F}\right)^{\mu\nu}Sp F^2]\right\}, \ea
 \ba
\nonumber && \rho^{\mu\nu}(\tau) =
C_{0}(\tau)\left\{2C_{1}(\tau)\left(\tilde{F}\frac{\sin^2F\tau}{F}\right)^{\mu\nu}
-2C_{2}(\tau)\left(\tilde{F}^{\mu\nu}\frac{\sin^2F\tau}{F}\right)^{\mu\nu}Sp
F\cot F\tau \right.- \\ \nonumber &-&
C_{3}(\tau)[2(\cos^2F\tau)^{\mu\nu} -
\frac{1}{2}\left(\frac{\sin2F\tau}{F}\right)^{\mu\nu}Sp F\cot
F\tau] - C_{4}(\tau)[4(F\sin2F\tau)^{\mu\nu} - \\ && \left.
 - 2(\sin^2F\tau)^{\mu\nu}Sp F\cot F\tau -
\left(\frac{\sin2F\tau}{F}\right)^{\mu\nu}Sp F^2]\right\}, \ea
 \ba
\nonumber && \omega^{\mu\nu}_{5}(\tau) =
C_{0}(\tau)\left\{C_{1}(\tau)[2(\cos^2 F\tau)^{\mu\nu} -
\frac{1}{2}\left(\frac{\sin 2F\tau}{F}\right)^{\mu\nu}Sp F\cot
F\tau] + \right.\\ \nonumber && + C_{2}(\tau)\left[4(F\sin
2F\tau)^{\mu\nu}- 2(\sin^2 F\tau)^{\mu\nu}Sp F\cot F\tau -
\left(\frac{\sin 2F\tau}{F}\right)^{\mu\nu}Sp F^2\right] + \\ &&
\left. + 2C_{3}(\tau)\left(\tilde{F}\frac{\sin^2
F\tau}{F}\right)^{\mu\nu} -
2C_{4}(\tau)\left(\tilde{F}\frac{\sin^2
F\tau}{F}\right)^{\mu\nu}Sp F\cot(F\tau)\right\}, \ea
 \ba && \nonumber
\rho^{\mu\nu}_{5}(\tau) = -C_{0}(\tau)\left\{C_{1}(\tau)[(\sin 2F\tau)^{\mu\nu}
- \left(\frac{\sin^2 F\tau}{F}\right)^{\mu\nu} Sp F\cot F\tau] - \right.\\  &-&
C_{2}(\tau)[4(F\cos 2F\tau)^{\mu\nu} -  (\sin 2F\tau)^{\mu\nu}Sp F\cot F\tau +
2\left(\frac{\sin^2 F\tau}{F}\right)^{\mu\nu}Sp F^2]- \\
\nonumber && \left. -
 C_{3}(\tau)\left(\tilde{F}\frac{\sin 2F\tau}{F}\right)^{\mu\nu} -
C_{4}(\tau)[4\tilde{F}^{\mu\nu} - \left(\tilde{F}\frac{\sin
2F\tau}{F}\right)^{\mu\nu}Sp F\cot F\tau]\right\}, \ea
 and notation
 \be C_{0}(\tau)=\tau^{-2}\exp\left(-\frac{1}{2}Sp \ln\frac{\sin
 F\tau}{F\tau}\right)\ee
  is introduced for brevity. Coefficient functions $C_{j}(\tau)$
  ($j=\overline{1,4}$) are given explicitly by expressions
\ba  \nonumber C_{1}(\tau) &=& Re
\cosh\left[\tau\sqrt{\frac{1}{2}(-SpF^2+\imath
SpF\tilde{F})}\right],
\\ \nonumber C_{2}(\tau) &=& Re \frac{\sinh\left[\tau\sqrt{\frac{1}{2}(-SpF^2+\imath
SpF\tilde{F})}\right]}{\sqrt{2(-SpF^2+\imath SpF\tilde{F})}},\\
 C_{3}(\tau) &=& Im
\cosh\left[\tau\sqrt{\frac{1}{2}(-SpF^2+\imath SpF\tilde{F})}\right],\\
\nonumber C_{4}(\tau) &=& Im
\frac{\sinh\left[\tau\sqrt{\frac{1}{2}(-SpF^2+\imath
SpF\tilde{F})}\right]}{\sqrt{2(-SpF^2+\imath SpF\tilde{F})}},\ea
 and $C_{0}(\tau)$ (4.12) is expressed as (see Ref.\cite{Sch})
 \be C_{0}(\tau)= SpF\tilde{F}[4C_{3}(\tau)]^{-1}. \ee
 There are remarkable relations among $C_{j}(\tau)$:
 \be C_{1}(\tau)(\tan F\tau)^{\mu\nu} = 2[C_{2}(\tau)F^{\mu\nu} +
 C_{4}(\tau)\tilde{F}^{\mu\nu}]\ee
  and
  \be C_{3}(\tau)(\cot F\tau)^{\mu\nu} = 2[C_{2}(\tau)\tilde{F}^{\mu\nu} -
 C_{4}(\tau)F^{\mu\nu}].\ee
 In Appendix we prove these relations  and find out that
  \be \omega^{\mu\nu}(\tau)=\rho^{\mu\nu}(\tau) =
  \rho^{\mu\nu}_{5}(\tau)=0,\ee
   and
  \ba \nonumber \omega^{\mu\nu}_{5}(\tau) &=& - g^{\mu\nu}\{Sp F^2 + \frac{[Sp
 F^4 - \frac{1}{2}(Sp F^2)^2](Sp F\cot F\tau)^2 + 4(Sp F^3\cot
   F\tau)^2}{2[4SpF^4 - (SpF^2)^2]}\} +\\ &+& (F^2)^{\mu\nu} \left[2 -
   \frac{Sp F\cot F\tau(SpF^2 Sp F\cot F\tau - 4 Sp F^3\cot
   F\tau)}{4SpF^4 - (SpF^2)^2}\right]. \ea

Thus, vector current $J^{\mu}$ (2.9) is vanishing, whereas
axial-vector current $J^{\mu5}$ (2.10) is real (as it should be)
and divergent. Regularizing the divergence by introducing a small
positive $\tau_{0}$ as the lower limit of the $\tau$-integral in
Eq.(4.7), and separating the divergent at $\tau_{0}\rightarrow 0$
term from the convergent one, we get
 \be J^{\mu5} = -\frac{b^{\mu}}{2\pi^2}\left[\frac{1}{\tau_{0}} +
 m^2\ln(m^2\tau_{0}e^{\gamma -1})\right] +
 \frac{1}{(2\pi)^2}\int^{\infty}_{0}d\tau e^{-\tau m^2}\left[\omega^{\mu\nu}_{5}(\tau) +
 \frac{2}{\tau^2}g^{\mu\nu}\right]b_{\nu}, \ee
  where $\gamma$ is the Euler constant.

\section{Effective lagrangian}

\hspace{0.7cm}Using either (3.20a) or (3.20b), we get
 \be <x(-\imath\tau)|x(0)> =
\frac{\imath}{(4\pi)^2}\frac{1}{\tau^2}\exp\left[-\tau m^2
-\frac{1}{2}Sp\ln\frac{\sin(F\tau)}{F\tau}\right]\exp\left(\frac{\tau}{2}Sp\sigma
F\right), \ee
 and, consequently,
 \be \frac{\imath}{2}\int^{\infty}_{0}\frac{d\tau}{\tau}tr<x|e^{-\tau\mathcal{H}}|x> =
  -\frac{1}{8\pi^2}\int^{\infty}_{0}\frac{d\tau}{\tau}e^{-\tau
  m^2}C_{0}(\tau)C_{1}(\tau),\ee
  where $C_{0}(\tau)$ and $C_{1}(\tau)$ are given by Eqs.(4.13) and
 (4.14). This coincides with the Schwinger's result \cite{Sch}, proving that the
  linear  in $b$  corrections to the effective action are absent, which is consistent with the
  vanishing of vector current $J^{\mu}$. The nonvanishing of
  axial-vector current $J^{\mu5}$ results in the appearance of the
  quadratic in $b$ corrections to the effective action.

  Really, Eq.(2.13), rewritten in terms of the effective
  lagrangian (density of the effective action), takes form
  \be \mathcal{L}^{eff}(A, b) - \mathcal{L}^{eff}(A, 0) =
  \frac{1}{2}b_{\mu}J^{\mu5}, \ee
  where we have used the linearity  of $J^{\mu5}$ (4.19) in $b$ and integrated over parameter $u$.
  Thus, identifying $\mathcal{L}^{eff}(A, 0)$ with Eq.(5.2) we get
   \be \mathcal{L}^{eff}(A, b) = -\frac{1}{8\pi^2}\int^{\infty}_{0}d\tau
   e^{-\tau m^2}[\frac{1}{\tau}C_{0}(\tau)C_{1}(\tau) -
   b\omega_{5}(\tau)b], \ee
   where $\omega^{\mu\nu}_{5}(\tau)$ is given by Eq.(4.18) and the first term in
   square brackets can be also
   presented in a similar manner as Eq.(4.18):
   \be C_{0}(\tau)C_{1}(\tau) = \frac{1}{16}\left[(Sp F\cot F\tau)^2 - \frac{(SpF^2 SpF\cot
   F\tau - 4 SpF^3\cot F\tau)^2}{4Sp F^4 - (Sp F^2)^2}\right]. \ee
   Regularizing divergence of the integral in Eq.(5.4) by
   introducing cut off $\tau_{0}$ and separating the divergent at
   $\tau_{0}\rightarrow 0$ terms from the convergent ones, we get
   \ba \nonumber && \mathcal{L}^{eff}(A, b) = \frac{1}{(4\pi)^2}\left[\frac{1}{\tau_{0}^2}-
   2\frac{m^2}{\tau_{0}} - m^4\ln(m^2\tau_0 e^{\gamma-\frac{3}{2}})\right]-
   \frac{Sp F^2}{3(4\pi)^2}\ln(m^2\tau_0 e^{\gamma}) -
   \\ &-& \frac{b^2}{(2\pi)^2}\left[\frac{1}{\tau_0} + m^2\ln(m^2\tau_0 e^{\gamma -
   1})\right]- \frac{1}{8\pi^2}\int_{0}^{\infty}d\tau e^{-\tau m^2}
   \left[\frac{1}{\tau}C_{0}(\tau)C_{1}(\tau) - \right.\\ && \left. -
   \nonumber \frac{1}{\tau^3} + \frac{Sp F^2}{6\tau} - b\omega_{5}(\tau)b -
   2\frac{b^2}{\tau^2}\right]. \ea
 Subtraction of terms in the first square brackets in the right
 hand side of Eq.(5.6) corresponds to the requirement that
 the effective lagrangian should vanish at vanishing $F$ and $b$.
 Subtraction of other terms which are not included into the
 convergent integral corresponds to the redefinition
 (renormalization) of bare parameters of the lagrangian of the
 bosonic sector. Namely, the logarithmically divergent term which is proportional to $Sp F^2$
 is combined with the Maxwell term to
 produce  the charge  renormalization: $\frac{Sp F^2}{4 e^2} \rightarrow \frac{Sp F^2}{4
 e^2_{ren}}$ \cite{Sch}. In a quite  similar way, the terms in the
 second square brackets in the right hand side  of Eq.(5.6) are
 absorbed into the renormalization of the coefficient before
 $b^2$: $-\kappa^2b^2\rightarrow -\kappa^2_{ren}b^2$. Thus, we are
 left with the finite renormalized effective lagrangian,
\be  \mathcal{L}^{eff}_{ren}(A, b) = - \frac{1}{8\pi^2}\int_{0}^{\infty}d\tau
e^{-\tau m^2}\left[\frac{1}{\tau}C_{0}(\tau)C_{1}(\tau) - \frac{1}{\tau^3} +
\frac{Sp F^2}{6\tau}  - b\omega_{5}(\tau)b -
   2\frac{b^2}{\tau^2}\right], \ee
 which with the use of Eqs.(4.18) and (5.5) is rewritten in explicit form
\ba && \nonumber \mathcal{L}^{eff}_{ren}(A, b) = -
\frac{1}{8\pi^2}\int_{0}^{\infty}d\tau e^{-\tau
m^2}\left(\frac{1}{16\tau}\left[(Sp F\cot F\tau)^2 -\right.\right.\\
&& \nonumber -\left. \frac{(Sp F^2 SpF\cot F\tau - 4 Sp F^3\cot
F\tau)^2}{ 4SpF^4 -
(Sp F^2)^2}\right] - \frac{1}{\tau^3} + \frac{Sp F^2}{6\tau} + \\
&&  +
 b^2\left\{SpF^2 + \frac{[Sp F^4 - \frac{1}{2}(SpF^2)^2](SpF\cot
F\tau)^2 + 4(Sp F^3\cot F\tau)^2}{2[4SpF^4 - (SpF^2)^2]}
 - \frac{2}{\tau^2}\right\} -\\\nonumber && \left.-  bF^2b\left[2
 -
\frac{Sp F\cot F\tau(Sp F^2 SpF\cot F\tau - 4SpF^3\cot
F\tau)}{4SpF^4 - (SpF^2)^2}\right]\right). \ea

 In the case of weak field strength, $F^{\mu\nu}\ll m^2$, the last
 expression takes form
 \be \mathcal{L}^{eff}_{ren}(A, b) =
 \frac{1}{12\pi^2}\left\{\frac{1}{120 m^4}\left[7SpF^4 -
 \frac{5}{2}(SpF^2)^2\right] + \frac{1}{m^2}\left(bF^2b -
 \frac{1}{2}b^2SpF^2\right)\right\}; \ee
 note that the terms of the zeroth order in $b$ were first
 obtained more than 65 years ago by Heisenberg and Euler
 \cite{Hei} and Weisskopf \cite{Wei}.

 In the case of purely electric or magnetic field one has $SpF^4 = \frac{1}{2}(Sp
 F^2)^2$ and $SpF^2 = 2E^2$ or $SpF^2 = -2H^2$, where $E$ and $H$
 are the absolute values of the electric and magnetic field
 strengths, correspondingly. Expression (5.8) takes form
 \ba && \nonumber \mathcal{L}^{eff}_{ren}(A, b) =
 -\frac{1}{8\pi^2}\int^{\infty}_{0}d\tau e^{-\tau m^2}\left\{\frac{1}{\tau}
 \left(\frac{E}{\tau}\cot E\tau - \frac{1}{\tau^2} + \frac{1}{3}E^2\right) -
 \right. \\ &&
  \left. - 2[\mathbf{b}^2 - E^{-2}(\mathbf{b}\mathbf{E})^2]
 \left(\frac{E^2}{\sin^2E\tau} - \frac{1}{\tau^2}\right)\right\}
 \ea
 in the case of purely electric field with strength $\mathbf{E}\hspace{0.3cm} (|\mathbf{E}| =
  E)$, and
\ba && \nonumber \mathcal{L}^{eff}_{ren}(A, b) =
 -\frac{1}{8\pi^2}\int^{\infty}_{0}d\tau e^{-\tau m^2}\left\{\frac{1}{\tau}
 \left(\frac{H}{\tau}\coth H\tau - \frac{1}{\tau^2} - \frac{1}{3}H^2\right) +
 \right. \\ && \left.
   + 2[(b^0)^2 - H^{-2}(\mathbf{b}\mathbf{H})^2]
 \left(\frac{H^2}{\sinh^2H\tau} - \frac{1}{\tau^2}\right)\right\}
 \ea
 in the case of purely magnetic field with strength $\mathbf{H} \hspace{0.3cm}(|\mathbf{H}| =
  H)$. Note that the effective lagrangian does not depend on the
  time component of $b^{\mu}$ in the case of purely electric
  field.

In the case of vector $\mathbf{E}$ directed along vector $\mathbf{H}$ one has
$SpF^{2n} = 2[E^{2n} + (-1)^n H^{2n}]$, and expression (5.8) takes form \ba  &&
\nonumber \mathcal{L}^{eff}_{ren}(A, b) =
-\frac{1}{8\pi^2}\int^{\infty}_{0}d\tau e^{-\tau m^2}\left\{\frac{1}{\tau}
 \left[EH \cot E\tau \coth H\tau - \frac{1}{\tau^2} + \frac{1}{3}(E^2 - H^2)\right]
+ \right. \\ && \left. + 2b^2\left(\frac{H^2}{\sinh^2H\tau} -
\frac{1}{\tau^2}\right) - 2 \left[\mathbf{b}^2 - \frac{(\mathbf{b}\mathbf{E})^2
+  (\mathbf{b}\mathbf{H})^2}{E^2 + H^2}\right]\left(\frac{E^2}{\sin^2E\tau} -
 \frac{H^2}{\sinh^2H\tau}\right)\right\}.
 \ea

At last, in the case of $E=H$, when $Sp F^2 = 0$ and $Sp F^4 =\frac{1}{4} (Sp
F\tilde{F})^2 = 4 H^4 \cos^2 \theta$ \hspace{0.3cm}($\theta$ is the angle
between vectors $\mathbf{E}$ and $\mathbf{H}$), expression (5.8) takes form \ba
&& \nonumber\mathcal{L}^{eff}_{ren}(A, b) =
 -\frac{1}{8\pi^2}\int^{\infty}_{0} d\tau e^{-\tau m^2}\left\{\frac{1}{\tau}
\left[H^2 \cos \theta \cot(\tau H \sqrt{\cos \theta})\coth(\tau H \sqrt{\cos
\theta}) - \frac{1}{\tau^2}\right] - \right. \\ \nonumber && \left. - 2b^2H^2
\left[\frac{\sin^2 \frac{\theta}{2}}{\sin^2(\tau H \sqrt{\cos \theta})} -
\frac{\cos^2 \frac{\theta}{2}}{\sinh^2(\tau H \sqrt{\cos \theta})} +
\frac{1}{\tau^2 H^2}\right] - \right. \\ && \left. - [2\mathbf{b}^2 H^2 -
(\mathbf{b}\mathbf{E})^2 - (\mathbf{b}\mathbf{H})^2 ]\left[\frac{1}{\sin^2(\tau
H \sqrt{\cos \theta})} - \frac{1}{\sinh^2(\tau H \sqrt{\cos
\theta})}\right]\right\}. \ea

\section{Conclusion}

\hspace{0.7cm}In the present paper we have used the proper time method
\cite{Fo, Sch} to calculate the effective action of the theory with lagrangian
(1.2) in the case when electromagnetic field strength $F^{\mu\nu}$ and vector
$b^{\mu}$ are space-time independent. Previous attempts to solve this task
\cite{MC, JC3, Si1} were unconvincing , because the dependence of
$\gamma$-matrices on the proper time had not been adequately taken into
account. Really, since the commutator of Hamilton operator $\mathcal{H}$ (2.14)
with $\sigma $ is nonzero, the latter has to evolve in proper time as well as
canonical variables, and the correct system of the evolution equations is given
by Eq.(3.11){\footnote{In the case of $b^{\mu}$ equal to zero, the evolution
equation for $\Sigma(s)$ decouples, and Hamilton operator (3.13) loses the
dependence on the evolution of $\Sigma(s)$, owing to relation $SpF\Sigma(s) =
SpF\Sigma(0)$ in this case.}}. We solve this system in the linear in $b$
approximation and get transition amplitude (3.20) and Green's function (3.22).
Further, we find that vector current $J^{\mu}$ (2.9) is vanishing, which
ensures that the Chern-Simons term is not induced, because, otherwise, the
current would be nonvanishing, $J^{\mu}=\frac{1}{2}\tilde{F}^{\mu\nu}k_{\nu}$,
see Eq.(1.1). Moreover, the vanishing of $J^{\mu}$ means that corrections to
the effective action of the first order in $b$ are absent, and parity is not
violated in this approximation, although it is explicitly violated in initial
lagrangian (1.2). Also, we find that axial-vector current $J^{\mu5}$ (2.10) is
nonvanishing and is given by gauge invariant expressions (4.18)--(4.19). This
allows us to get corrections to the effective action of the second order in
$b$,  and we find that the renormalized (finite) effective lagrangian is given
by Eq.(5.8). It should be noted that the $\tau$-integral for the quadratic in
$b$ terms in Eq.(5.8) is indeed convergent in the case of purely magnetic field
only, see Eq.(5.11). In the case of nonvanishing electric field, when the terms
of zeroth order in $b$ develop nonvanishing imaginary part due to simple poles
of the cotangent function, the $\tau$-integral for the quadratic in $b$ terms
is divergent due to double poles of the inverse squared sine function, see
Eqs.(5.10), (5.12) and (5.13). Thus, the latter $\tau$-integral is not to be
understood literally but, instead, regarded just as a mere algorithm to get
terms up to any finite order in powers of $F^2/m^4$; in particular, for the
lowest nonvanishing order, see Eq.(5.9).

It should be emphasized that vanishing of vector current $J^{\mu}$
is related to the use of the approximation of the space-time
independent field strength. If the field strength is
inhomogeneous, then the current is nonvanishing even in the zeroth
order in $b$. Whether the inhomogeneity of the field strength
results in linear in $b$ corrections to $J^{\mu}$ and,
consequently, in parity violating terms in the effective action,
remains to be an open question which needs further investigation.

\section*{Acknowledgements}

\hspace{0.7cm}The work was supported by INTAS (grant INTAS OPEN
00-00055) and Swiss National Science Foundation (grant SCOPES
2000-2003 7 IP 62607).

\section*{Appendix}
\renewcommand{\theequation}{A.\arabic{equation}}
\setcounter{equation}{0}

\hspace{0.7cm}Let us consider quantity
 \ba && \nonumber \lambda^{\mu\nu}(\tau) = \omega^{\mu\nu} + (\rho\cot
 F\tau)^{\mu\nu} = -C_{0}\{4C_{2}\tilde{F}^{\mu\nu} + C_{3}[2(\cot F\tau)^{\mu\nu} -
 (F^{-1})^{\mu\nu}SpF\cot
 F\tau]+ \\ && + 2 C_{4}[2F^{\mu\nu} -
 (F^{-1})^{\mu\nu}SpF^2]\}.\ea
  Since $\lambda^{\mu\nu}$ contains only odd powers of the field
  strength, see Eqs.(4.8) and (4.9), its most general form is
   \be \lambda^{\mu\nu}(\tau) = \Lambda_{1}(\tau)F^{\mu\nu} +
   \Lambda_{2}(\tau)\tilde{F}^{\mu\nu},\ee
   where
   \be \nonumber \Lambda_{1}(\tau) = \frac{4SpF^3\lambda(\tau) -
   SpF^2SpF\lambda(\tau)}{(SpF^2)^2+(SpF\tilde{F})^2},\ee
  \be \Lambda_{2}(\tau) = (SpF\tilde{F})^{-1}\left[SpF\lambda(\tau) -
  SpF^2\frac{4SpF^3\lambda(\tau) -
   SpF^2SpF\lambda(\tau)}{(SpF^2)^2+(SpF\tilde{F})^2}\right].\ee
   Multiplying Eq.(A.1) by appropriate powers of the field
   strength and taking traces, we find
    \ba  \nonumber SpF\lambda(\tau) &=& 2C_{0}(C_{3}SpF\cot F\tau - 2C_{2}SpF\tilde{F} +
   2C_{4}SpF^2), \\
    SpF^3\lambda(\tau) &=& -\frac{1}{2}C_{0}SpF\tilde{F}(C_{3}Sp\tilde{F}\cot
   F\tau+ 2C_{2}SpF^2 + 2C_{4}SpF\tilde{F}). \ea
   By using the eigenvalue method of Schwinger (see Ref. \cite{Sch}), one
   can express $SpF \cot F\tau$ and $Sp\tilde{F}\cot F\tau$ via
    $SpF^2$ and $Sp F\tilde{F}$. The eigenvalue equation for $F$ has
    four solutions with eigenvalues $\pm f^{(1)}$ and $\pm
   f^{(2)}$, where
    \ba \nonumber && f^{(1)} = \frac{\imath}{2\sqrt{2}}[(-SpF^2 +\imath SpF\tilde{F})^{\frac{1}{2}} +
   (-SpF^2 -\imath SpF\tilde{F})^{\frac{1}{2}}], \\ && f^{(2)} =
   \frac{\imath}{2\sqrt{2}}[(-SpF^2 +\imath
   SpF\tilde{F})^{\frac{1}{2}} -  (-SpF^2 -\imath
   SpF\tilde{F})^{\frac{1}{2}}],\ea
  and eigenvalues of $\tilde{F}$ are related to those of $F$:
 \be \tilde{f}^{(l)} = -\frac{Sp F\tilde{F}}{4f^{(l)}}, \hspace{1cm} l=1, 2. \ee
Thus we get
   \be SpF\cot F\tau = 2[f^{(1)}\cot (f^{(1)}\tau) + f^{(2)}\cot
   (f^{(2)}\tau)], \ee
and
 \be Sp\tilde{F}\cot F\tau = -\frac{ Sp F\tilde{F}}{2f^{(1)}f^{(2)}}[f^{(2)}\cot (f^{(1)}\tau)
 + f^{(1)}\cot (f^{(2)}\tau)], \ee
which, upon substitution of Eq.(A.5), yield \be SpF\cot F\tau =
\frac{2}{C_{3}}(C_{2}Sp F\tilde{F} - C_{4}SpF^2) \ee
 and
  \be Sp\tilde{F}\cot F\tau =  -\frac{2}{C_{3}}(C_{2}Sp F^2 + C_{4}SpF\tilde{F}).\ee
 The last relations ensure that traces in Eq.(A.4) are equal to zero, and
 consequently,
 \be \lambda^{\mu\nu}(\tau) = 0. \ee
  Now, using Eq.(A.9), we can get rid of terms $(F^{-1})^{\mu\nu}$
  in Eq.(A.1), and get relation (4.16) in Section 4.

  Let us consider quantity
  \ba && \nonumber \lambda^{\mu\nu}_{5}(\tau) =
  \omega^{\mu\nu}_{5} + (\rho_{5}\cot F\tau)^{\mu\nu}
 = 2C_{0}\left\{C_{2}[2(F\cot F\tau)^{\mu\nu} - g^{\mu\nu} Sp F\cot
  F\tau]+ C_{3}(\tilde{F}F^{-1})^{\mu\nu} + \right.\\ \nonumber &&
  \left.
  + C_{4}[2(\tilde{F}\cot F\tau)^{\mu\nu} - (\tilde{F}F^{-1})^{\mu\nu}Sp F\cot
  F\tau]\right\} =  \frac{1}{2}\left\{\frac{C_{2}}{C_{3}}SpF \tilde{F}[2(F\cot F\tau)^{\mu\nu} -
 \right. \\ && \left. - g^{\mu\nu}SpF\cot F\tau] + 4(\tilde{F}^2)^{\mu\nu} + \frac{C_{4}}{C_{3}}[2(F\cot F\tau)^{\mu\nu}SpF\tilde{F} - 4(\tilde{F}^2)^{\mu\nu}Sp F\cot
  F\tau]\right\}, \ea
  where in the last line Eq.(4.14) is used. Since
  $\lambda^{\mu\nu}_{5}$contains only even  powers of the field
  strength, see Eqs.(4.10) and (4.11), its most general form is
  \be \lambda^{\mu\nu}_{5}(\tau) =
  \Omega_{1}(\tau)g^{\mu\nu}+
  \Omega_{2}(\tau)(\tilde{F}^2)^{\mu\nu}=  [\Omega_{1}(\tau) -
  \frac{1}{2}\Omega_{2}(\tau)SpF^2]g^{\mu\nu} +
  \Omega_{2}(\tau)(F^2)^{\mu\nu}, \ee
  where, due to the first relation in Eq.(4.5), either
  $(\tilde{F}^2)^{\mu\nu}$ or $(F^2)^{\mu\nu}$ can be chosen as
  complimentary to $g^{\mu\nu}$. Similarly to Eq.(A.3), scalar
  functions in Eq.(A.13) are related to appropriate traces:
  \ba && \nonumber \Omega_{1}(\tau) = \frac{1}{4}Sp\lambda_{5}(\tau) +
  \frac{1}{4}SpF^2\frac{4SpF^2\lambda_{5}(\tau) - SpF^2Sp\lambda_{5}(\tau)}{(SpF^2)^2 +
  (SpF\tilde{F})^2}, \\ && \Omega_{2}(\tau) = \frac{4SpF^2\lambda_{5}(\tau) -
  SpF^2Sp\lambda_{5}(\tau)}{(SpF^2)^2 +
  (SpF\tilde{F})^2}. \ea
  Using Eqs.(A.9), (A.10) and first two relations in Eq.(4.5), we
  get
  \ba &&\Omega_{1}(\tau) = -8C_{0}^{2}(C_{2}^{2} + C_{4}^{2}) =
  -\frac{(SpF\tilde{F})^2[(SpF\cot F\tau)^2 + (Sp\tilde{F}\cot
  F\tau)^2]}{8[(SpF^2)^2 + (SpF\tilde{F})^2]}, \ea
 and
 \ba && \nonumber \Omega_{2}(\tau) = 2[1- 2C_{3}^{-2}C_{4}(C_{2}SpF\tilde{F} -
 C_{4}SpF^2)] = \\ && = 2 + Sp F\cot F\tau\frac{SpF^2SpF\cot F\tau + SpF\tilde{F}Sp\tilde{F}\cot
 F\tau}{(SpF^2)^2 + (SpF\tilde{F})^2}. \ea
\hspace{1cm} Using Eqs.(A.9) and (4.16), we reduce Eq.(4.9) to the
form
 \ba && \rho^{\mu\nu}(\tau) = 2C_{0}[C_{1}g^{\mu}_{\beta} - 2(C_{2}F^{\mu\alpha} + C_{4}\tilde{F}^{\mu\alpha})(\cot
 F\tau)_{\alpha\beta}]\left(\tilde{F}\frac{\sin^2
 F\tau}{F}\right)^{\beta\nu}, \ea
 and, similarly, Eq.(4.11) with the use of Eq.(4.16) is reduced to
 the form
 \ba \nonumber && \rho^{\mu\nu}_{5}(\tau) = -C_{0}[C_{1}g^{\mu}_{\beta} - 2(C_{2}F^{\mu\alpha} + C_{4}\tilde{F}^{\mu\alpha})(\cot
 F\tau)_{\alpha\beta}]\times \\&& \times[(\sin 2F\tau)^{\beta\nu} - (\frac{\sin^2
 F\tau}{F})^{\beta\nu}SpF\cot F\tau]. \ea
 Thus, in order to prove the vanishing of $\rho^{\mu\nu}$ and
 $\rho_{5}^{\mu\nu}$, it is sufficient to prove relation
 \be C_{1}g^{\mu}_{\beta} - 2(C_{2}F^{\mu\alpha} +
 C_{4}\tilde{F}^{\mu\alpha})(\cot F\tau)_{\alpha\beta} = 0. \ee
 First, using again Eq.(4.16), we get
 \be (C_{2}F^{\mu\alpha} + C_{4}\tilde{F}^{\mu\alpha})(\cot
 F\tau)_{\alpha\beta} = g^{\mu}_{\beta}(2C_{3})^{-1}[(C_{2}^{2} - C_{4}^{2})SpF\tilde{F} -
 2C_{2}C_{4}SpF^2]. \ee
 Then, using explicit form of $C_{j}$ (4.13), we find relation
 \be C_{1}C_{3} = (C_{2}^{2} - C_{4}^{2})SpF\tilde{F} -
 2C_{2}C_{4}SpF^2, \ee
 which, together with the previous relation, proves Eq.(A.19) and, consequently,
 Eq.(4.15).

 Thus,  $\rho^{\mu\nu}$ and $\rho^{\mu\nu}_{5}$ are equal to zero,
 and, in view of Eq.(A.11), $\omega^{\mu\nu}$ is also equal to zero,
 whereas $\omega^{\mu\nu}_{5}$ is equal to
 $\lambda^{\mu\nu}_{5}$ (A.13), which with the use of Eq.(4.5) is
 recast into the form with dual field $\tilde{F}$ eliminated,
 resulting in Eq.(4.18).


\begin{thebibliography}{99}
\bibitem{Ca}
S.M. Carroll, G.B. Field, and R. Jackiw, Phys. Rev. D41 (1990)
1231.
\bibitem{Co1} D. Colladay and V.A. Kostelecky, Phys. Rev.
D55 (1997) 6760; D58 (1998) 116002.
\bibitem{CG}
S. Coleman and S.L. Glashow, Phys. Rev.  D59 (1999) 116008.
\bibitem{Ja1}
R. Jackiw, Comm. Mod. Phys.  A1 (1999) 1.
\bibitem{Kost}
CPT and Lorentz Symmetry, V.A. Kostelecky, Editor (World
Scientific, Singapore, 1999).
\bibitem{Jao} R. Jackiw and S. Templeton, Phys. Rev.  D23
(1981) 2291.
\bibitem{Go}
M. Goldhaber and V. Trimble, J. Astrophys. Astron.  17 (1996) 17;
S.M. Carroll and G.B. Field, Phys. Rev. Lett.  79 (1997) 2394.
\bibitem{Ja2}
R. Jackiw and V.A. Kostelecky, Phys. Rev. Lett.  82 (1999) 3572.
\bibitem{WC}
W.F. Chen, Phys. Rev.  D60 (1999) 085007.
\bibitem{PV}
M. Perez-Victoria, Phys. Rev. Lett.  83 (1999) 2518.
\bibitem{JC1}
J.-M. Chung and P. Oh, Phys. Rev.  D60 (1999) 067702; J.-M. Chung,
Phys. Lett.  B461 (1999) 138.
\bibitem{And}
A.A. Andrianov, P. Giacconi, and R. Soldati, JHEP 0202 (2002) 030.
\bibitem{Ja3}
R. Jackiw, Intern. J. Mod. Phys.  14 (2000) 2011.
\bibitem{Pe}
M. Perez-Victoria, JHEP 0104 (2001) 032.
\bibitem{Ba}
O.A. Battistel and G. Dallabona, Nucl. Phys. B610 (2001) 316.
\bibitem{JC2}
J.-M. Chung, Phys. Rev.  D60 (1999) 127901.
\bibitem{MC}
M. Chaichian, W.F. Chen, and R. Gonzalez Felipe,
Phys. Lett.  B503 (2001) 215.
\bibitem{JC3}
J.-M. Chung and  B.K. Chung, Phys.Rev. D63 (2001) 105015.
\bibitem{Si1}
Yu.A. Sitenko, Phys. Lett. B 515 (2001) 414.
\bibitem{Bo}
G. Bonneau, Nucl. Phys. B593 (2001) 398.
\bibitem{Ada}
C. Adam and F.R. Klinkhamer, Phys. Lett. B513 (2001) 245.
\bibitem{Fo}
V. Fock, Physik. Z. Sowjetun.  12 (1937) 404.
\bibitem{Sch}
J. Schwinger, Phys. Rev.  82 (1951) 664.
\bibitem{Sal}
A. Salam and J. Strathdee,  Nucl. Phys. B 90 (1975) 203.
\bibitem{Dow}
J.S. Dowker and R. Critchley, Phys. Rev. D 13 (1976) 3224 .
\bibitem{Haw}
S.W. Hawking, Commun. Math. Phys. 55 (1977) 133 .
\bibitem{Hei}
W. Heisenberg and H. Euler, Z. Phys. 98 (1936) 714.
\bibitem{Wei}
V. Weisskopf, Kgl. Dan. Vid. Selsk. Mat. Fys. Medd. 14 (1936) 6.


\end{thebibliography}
\end{document}